\documentclass[letterpaper, 10 pt, conference]{ieeeconf}  

\IEEEoverridecommandlockouts                              

\title{\LARGE \bf
A Subject-Independent Brain-Computer Interface Framework Based on Supervised Autoencoder
}

\author{Navid Ayoobi$^{*}$ and Elnaz Banan Sadeghian$^{*}$
\thanks{$^{*}$Navid Ayoobi and Elnaz Banan Sadeghian are with the Department of Electrical and Computer Engineering,
        Stevens Institute of Technology, Hoboken, USA.
        {(E-mail: \tt\small  nayoobi@stevens.edu,}
        {\tt\small ebsadegh@stevens.edu})
        }%
}

\usepackage{cite}
\usepackage{amsmath,amssymb,amsfonts}
\usepackage{algorithmic}
\usepackage{graphicx}
\usepackage{textcomp}
\usepackage{longtable}
\usepackage{multirow}
\usepackage{xcolor}
\usepackage{color, colortbl}
\definecolor{Gray}{gray}{0.92}
\definecolor{DarkGray}{gray}{0.8}
\usepackage{caption}
\usepackage{float}
\usepackage{diagbox}

\def\BibTeX{{\rm B\kern-.05em{\sc i\kern-.025em b}\kern-.08em
    T\kern-.1667em\lower.7ex\hbox{E}\kern-.125emX}}

\begin{document}

\maketitle
\thispagestyle{empty}
\pagestyle{empty}

\begin{abstract}

A calibration procedure is required in motor imagery-based brain-computer interface (MI-BCI) to tune the system for new users.
This procedure is time-consuming and prevents naive users from using the system immediately.
Developing a subject-independent MI-BCI system to reduce the calibration phase is still challenging due to the subject-dependent characteristics of the MI signals.
Many algorithms based on machine learning and deep learning have been developed to extract high-level features from the MI signals to improve the subject-to-subject generalization of a BCI system.
However, these methods are based on supervised learning and extract features useful for discriminating various MI signals.
Hence, these approaches cannot find the common underlying patterns in the MI signals and their generalization level is limited.
This paper proposes a subject-independent MI-BCI based on a supervised autoencoder (SAE) to circumvent the calibration phase.
The suggested framework is validated on dataset 2a from BCI competition IV.
The simulation results show that our SISAE model outperforms the conventional and widely used BCI algorithms, common spatial and filter bank common spatial patterns, in terms of the mean Kappa value, in eight out of nine subjects.

\end{abstract}

\begin{figure*}[h]
    \centering
    \includegraphics[width=1.3\columnwidth]{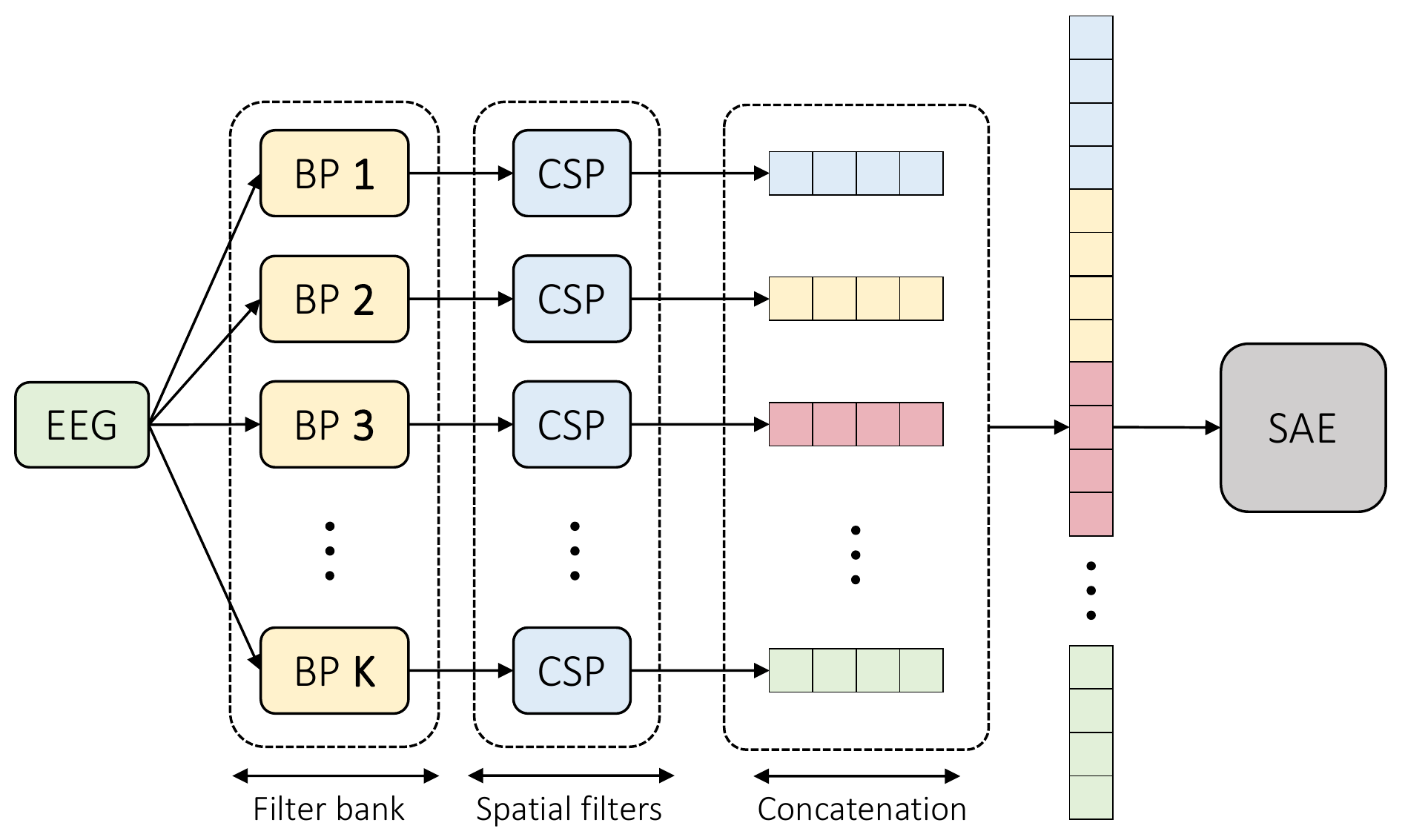}
    \caption{\small The procedure of extracting features. The EEG signals are bandpass filtered using $K$ different frequency ranges represented in $\mathcal{F}$. The CSP algorithm is applied to the filtered signals to generate a feature vector. These vectors are then fused to feed a supervised autoencoder.}
    \label{fig:FE}
\end{figure*}
\section{INTRODUCTION}

A brain-computer interface (BCI) is a system that directly links brain activities to external devices in order to enable people with movement disabilities \cite{zhou2020hybrid,sadeghian2008fractal}. 
Motor imagery electroencephalography (MI-EEG) is a non-invasive technique used in BCI to acquire brain activities after rehearsing a motor act. 
Generally, an MI-BCI system is ready to use after a calibration procedure.
The calibration includes acquiring MI-EEG signals from a subject and training a model on collected data.
It takes approximately $20{-}30$ minutes to complete this procedure \cite{blankertz2007non}.
Since some characteristics of EEG signals, for example the spatial origin of the signals, vary from one subject to another, a new calibration procedure is required for each new user.
As a result, the instant use of a BCI system is not possible for each new user.
Nevertheless, most conventional BCI studies are dedicated to designing a system based on subject-dependent approaches \cite{koles1990spatial,ang2012filter}.
These approaches still need calibration to be generalized to a new subject.

To alleviate the subject-dependency issue, BCI researchers aim to reduce the time or the number of training samples required for the calibration phase by leveraging data collected from other individuals \cite{jayaram2016transfer,jiao2018sparse,lotte2010learning}: 
Jayaram \textit{et al.} \cite{jayaram2016transfer} propose a framework based on transfer learning to reduce the training time needed in a subject-to-subject or session-to-session transfer in an MI paradigm.
In order to decrease the required training samples for one subject, Jiao \textit{et al.} \cite{jiao2018sparse} establish a sparse group representation model to find the most compact representation of a test sample based on a linear combination of the common spatial pattern (CSP) features extracted from training samples of all available subjects.
However, minimum data must still be acquired from new subjects in these approaches.
Therefore, a naive user is still unable to utilize the designed BCI system immediately.
On the other hand, zero-calibration approaches attempt to eliminate the calibration phase in order to ready a BCI system for instant usage by inexperienced users \cite{lotte2009comparison,joadder2019new}:
Lotte \textit{et al.} \cite{lotte2009comparison} develop a subject-independent (SI) method utilizing a multi-resolution frequency decomposition algorithm for finding the most generalizable frequency ranges in filter bank CSP (FBCSP).
Joadder \textit{et al.} \cite{joadder2019new} find common discriminating patterns among different subjects by exploiting four different feature extraction methods. These features were then fed to a linear discriminant analysis (LDA) classifier in their proposed SI-BCI method.
Nevertheless, most of the above zero-calibration methods rely only on the linear characteristics of the EEG signals.

In recent years, machine learning and deep learning have shown promising results in automatically extracting distinguishable features from EEG signals through non-linear processes \cite{kwon2019subject,zhang2019convolutional}: 
Kwon \textit{et al.} \cite{kwon2019subject} construct a large MI-EEG dataset and introduce an SI-BCI framework based on the deep convolutional neural network (CNN).
However, current methods use supervised learning and extract features that accurately map the input data onto labels. 
Hence, the trained model cannot find the common underlying representation of different subjects.
This fact results in a poor generalization to other datasets.

In this paper, we propose a zero-calibration method to develop a BCI system for immediate use.
We utilize a large filter bank to extract features from MI-EEG signals and feed them to our subject-independent supervised autoencoder (SISAE).
The autoencoder within the SISAE extracts non-linear features representing the underlying patterns of the EEG signals.
The classifier of the SISAE forces the autoencoder to extract those underlying features that are suitable for discriminating the desired MI signals. 
Therefore, the SISAE makes a trade-off between finding the common underlying patterns and the features suited for classification.

To evaluate the generalization performance of the proposed SISAE, we utilize dataset 2a from the BCI competition IV, which consists of nine subjects.
For each subject, we train the SISAE using the other eight subjects.
The simulation results show that the suggested method can extract the common underlying patterns of MI-EEG signals among different subjects and provide a promising generalization performance.
The SISAE outperforms the CSP and FBCSP algorithm in eight out of nine subjects in terms of the mean kappa value.

The remainder of this article is organized as follows.
In section \ref{dataset}, we describe the dataset.
The proposed method is elaborated in section \ref{method}. 
The results are presented and discussed in section \ref{results}. 
Section \ref{conclusion} concludes the article.

\section{Dataset} \label{dataset}

We use dataset 2a from the BCI competition IV \cite{brunner2008bci}. 
In this dataset, $22$ electrodes are used to collect EEG signals from nine subjects. 
The subjects performed four motor imageries: the left hand, the right hand, the feet and the tongue.
The training and testing datasets are recorded on different days. 
Each dataset contains $72$ trials for each class.
No feedback is provided during the recording.
The recording procedure for each trial starts with a warning tone and a fixation cross on the screen. 
At $t{=}2s$, an arrow appears on the screen for $1.25s$ to ask the subject to perform the motor imagery until $t{=}6s$.
For this paper, we only use the signals of the left and right hands for our binary classification.
We also extract the interval from second $0.5$ to the second $2.5$ of the recorded trials for our processing, similar to the procedure in \cite{ang2012filter}.

\section{Methods} \label{method}

The spectral and spatial information in the MI signals are subject-dependent.
In a subject-specific method, the most discriminative frequency bands and spatial regions are identified for each subject to enhance the system performance.
However, in designing a subject-independent framework, the challenge is to extract features that can be better generalized to other subjects. 
To this end, we employ a large filter bank and apply CSP algorithm \cite{koles1990spatial} to extract the spatial patterns of each bandpass filtered signal.
The obtained features in different frequency bands are fused to feed the proposed subject-independent supervised autoencoder (SISAE) network explained in \ref{SISAE}.

\subsection{Feature extraction}

We define a large set of frequency bands in $\mathcal{F}$ to form our filter bank.
The set $\mathcal{F}$ covers the frequencies between $4$ Hz to $40$ Hz and includes the frequency bands with bandwidth changing from $1$ Hz to $36$ Hz according to
\begin{equation}
    \mathcal{F} = \Big\{[4,5],[5,6],...,[5,40],[4,40]\Big\}.
\end{equation}
Each EEG signal is accordingly bandpass filtered with a sixth-order Butterworth filter with cutoff frequencies given in the $i$-th frequency band $\mathcal{F}_i$ in the set $\mathcal{F}$.
The signals filtered with $\mathcal{F}_i$ are fed to the CSP algorithm with $m$ pairs of spatial filters to produce a feature vector $\mathbf{V}_i$.
The obtained vectors in different frequency bands are concatenated to form a larger feature vector $\mathbf{V}$ with a size of $2mK$ where $K$ is the number of frequency bands represented in $\mathcal{F}$.
This procedure is illustrated in Fig. \ref{fig:FE}.

\subsection{Subject-independent supervised autoencoder (SISAE)}\label{SISAE}

In supervised learning, the neural network does not necessarily learn the underlying patterns in the data so that it suffers from the generalization issue \cite{le2018supervised}. 
On the other hand, unsupervised learning strategies may not be effective in classifying different MI tasks.
In this article, we propose a network that jointly learns the supervised tasks, here, the classification of the left versus right hand, and the underlying patterns for better generalization.

\begin{figure}[t!]
    \centering
    \includegraphics[width=0.6\columnwidth]{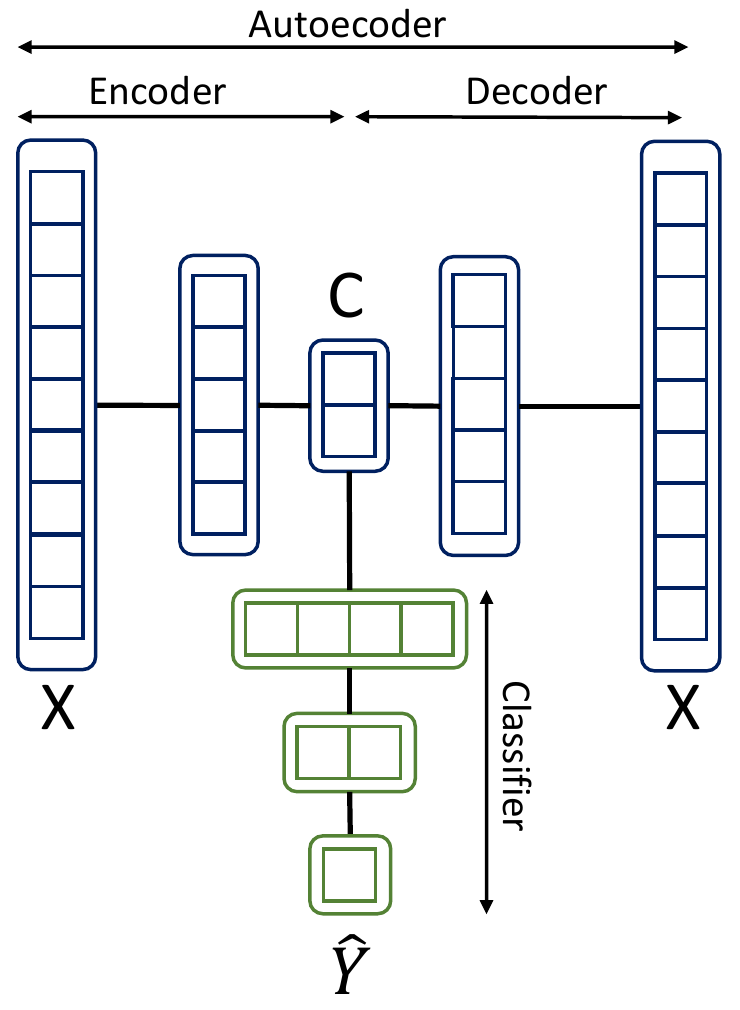}
    \caption{\small Proposed subject-independent supervised autoencoder (SISAE).}
    \label{fig:SAE}
\end{figure}

The proposed SISAE architecture is depicted in Fig. \ref{fig:SAE}.
It is composed of an autoencoder network and a fully connected feed-forward binary classifier. 
The AE learns the underlying representation of the data by reconstructing the input.
The encoder maps the input onto a code vector $\mathbf{C}{=}Enc(\mathbf{X})$.
The decoder takes the code vector and reconstructs the input $\mathbf{X}{=}Dec(\mathbf{C})$.
To prevent the AE from copying the input, the latent layer's dimensionality is set to a number smaller than the input dimensionality.
The classifier is then fed with $\mathbf{C}$.
Both networks are trained simultaneously to minimize a composite loss function $\mathbf{Q}$.
The $\mathbf{Q}$ comprises a reconstruction loss $\mathbf{Q_r}$ and a loss for classification task $\mathbf{Q_c}$ as follows

\begin{equation}
\mathbf{Q} = \frac{1}{N} \sum_{n=1}^{N}\Big(\alpha\mathbf{Q_c}(W_cW_ex_i,y_i)+\beta \mathbf{Q_r}(W_dW_ex_i,x_i) \Big),
\end{equation}
where $N$, $W_e$, $W_d$, $W_c$, $x_i$ and $y_i$ are the number of trials in the training set, the weights of the encoder, the weights of the decoder, the weights of the classifier, the $i$-th input and its corresponding label, respectively. 
The hyperparameters $\alpha$ and $\beta$ are the loss weights that are tuned in cross validation.
We define the reconstruction loss $\mathbf{Q_r}$ as the mean squared error
\begin{equation}
    \mathbf{Q_r}(W_dW_ex_i,x_i) = \frac{1}{|x_i|} \|W_dW_ex_i - x_i \|^2 ,
\end{equation}
where $|x_i|$ is the input dimensionality. The classification loss $\mathbf{Q_c}$ is defined as a binary cross entropy loss
\begin{equation}
    \mathbf{Q_c}(W_cW_ex_i,y_i) = -\big ( y_i P(y_i) + (1-y_i)P(1-y_i) \big ),
\end{equation}
where $P(.)$ is the predicted probability calculated by a sigmoid function as the activation function of the last layer in the classifier network. 

\section{Results and discussion}\label{results}
\subsection{Cross validation and parameter setting}

For training the SISAE network, we use eight training sets corresponding to eight out of nine subjects, excluding one subject for testing.
To avoid overfitting, we add an $L_1$ and an $L_2$ regularization terms to the loss function. Here, we set the regularization factors, learning rate, and the mini batch to $0.0001$, $0.01$, and $32$ for all the experiments.
In order to prevent AE from overfitting, we divided the total number of epochs into $50$ and $150$ epochs, and simultaneously trained both the AE and the classifier during the first $50$ epochs, leaving the last $150$ epochs to only train the classifier while the AE weights are frozen.
\begin{table}[!t]
    \centering
    \caption{\small Different settings for LOSO cross-validation}
    \begin{tabular}{|c|cc|}
    \hline
    \rule{0pt}{1.em}Setting &  AE nodes & Classifier nodes\\
    \hline
     \rule{0pt}{1.2em}  1  & [5,3,5] & [3,3,3,1]\\[0.2em]
       2  & [10,5,10] & [5,5,5,1]\\[0.2em]
       3  & [20,10,20] & [10,5,5,1]\\[0.2em]
       4  & [30,15,30] & [15,10,5,1]\\[0.2em]
       5  & [40,20,40] & [15,10,5,1]\\[0.2em]
       
    \hline
    \end{tabular}
    \label{tab:configurations}
\end{table}

\begin{table}[t]
    \centering
    \caption{\small Cross validation results in terms of mean Kappa value}
    \setlength{\tabcolsep}{4.4pt}
    \begin{tabular}{|c|ccccc|cc|}
    
    \hline
     \backslashbox[1cm]{Sub.}{Sett.} & 1 & 2 & 3 & 4 & 5 & Mean & Std\\
    \hline
       \rule{0pt}{1.2em} 1  & 0.3534 &\textbf{0.3839} &0.3765 &0.3584 &0.3596&0.3664&0.0131 \\[0.2em]
       2  & 0.4615 &0.4658 &\textbf{0.4856}  &0.4733 &0.4715& 0.4715               &0.0091 \\[0.2em]
       3  & 0.4411 &0.4384 &0.4385 &\textbf{0.4429} &0.4392 & 0.4400               &0.0019 \\[0.2em]
       4  & \textbf{0.4901} &0.4855 &0.4882 &0.4697 &0.4666 & 0.4800               &0.0110 \\[0.2em]
       5  & 0.4808 &0.4886 &0.4929 &0.4875 &\textbf{0.4941} & 0.4888               &0.0053 \\[0.2em]
       6  & 0.4355 &0.4520 &0.4508 &0.4665 &\textbf{0.4676} & 0.4545               &0.0132 \\[0.2em]
       7  & 0.4452 &0.4499 &\textbf{0.4566} &0.4447 &0.4500 & 0.4493               &0.0048 \\[0.2em]
       8  & 0.3886 &0.3884 &0.3799 &\textbf{0.3950} &0.3937 & 0.3891               &0.0059 \\[0.2em]
       9  & 0.5034 &0.5046 &0.5035 &0.5104 &\textbf{ 0.5141}& 0.5072               &0.0048 \\[0.2em]
         \hline
    \end{tabular}
    \label{tab:crossvalidation}
\end{table}

To obtain the proper model parameters, we utilize the leave-one-subject-out (LOSO) strategy for cross validation \cite{ray2015subject}.
For example, assume that the test subject is subject 9.
We perform the cross validation on the remaining eight subjects.
We choose the training set of one of the eight subjects as the validation set and train the SISAE network on the remaining seven subjects.
This way, we train the SISAE network eight times for each specific test subject.

Table \ref{tab:configurations} shows different settings for hidden layers of AE and classifier.
The results of the cross validation for each of these configurations and each subject are presented in Table \ref{tab:crossvalidation} in terms of the mean kappa value \cite{cohen1960coefficient}.
The best Kappa value for each subject is shown in boldface.
According to the obtained standard deviation values, there is no significant difference between the system performances under various settings. 
Therefore the proposed model is robust with respect to the changes in the model structure.
Nevertheless, we chose the best setting for each subject. 
Moreover, based on the mean Kappa values, it is worth mentioning that some of the subjects provide more generalizing features, yielding better performance on other subjects. For instance, when subjects one and eight are the test subjects and therefore are removed from the training set, the averaged mean Kappa values across different settings are low and equal to $0.3664$ and $0.3891$, respectively.
As a result, at least in our experiment, a careful selection of good subjects helps to improve the generalization performance of our system.

\subsection{Comparison of SISAE with CSP and FBCSP methods}

We evaluate our proposed SISAE by comparing it with the CSP \cite{koles1990spatial} and FBCSP \cite{ang2012filter} algorithms.
For the CSP, the EEG signals are bandpass filtered between  $4$ HZ and $40$ Hz.
For the FBCSP, nine bandpass filters, covering the frequency range of $4{-}40$ Hz, are used and the mutual information-based best individual feature algorithm is utilized to select the spatial features.
For all methods, we used $m{=}2$ pairs of the spatial filters to extract the features.
In addition, an  LDA classifier is used to classify the spatial features extracted by the CSP and FBCSP algorithms.

Table \ref{tab:results} shows the mean Kappa value obtained for each subject.
We observe that the proposed method outperforms conventional methods in eight out of nine subjects.
Further, we observe the superiority of the proposed method for the subjects with low performance (Kappa$\,<0.1$) corresponding to the CSP and FBCSP methods.
The reason is that in the CSP and FBCSP the classifier is trained by directly mapping the subject-dependent features onto the labels and therefore it performs poorly on the new subject.
To the contrary, the autoencoder within the SISAE network extracts the underlying patterns and the classifier maps these patterns onto labels.
Further, we observe that the conventional methods perform nearly similar to a random classifier for the subjects $2$, $5$, $6$, and $7$ where our proposed method performs notably better.

The average Kappa value across all subjects are $0.226$, $0.218$, and $0.500$ for CSP, FBCSP, and SISAE, respectively.
The Kappa value improvement by our proposed SISAE is statistically significant.
The p-value of the paired t-test with a confidence interval of $95\%$ between the proposed SISAE and the two other methods is less than $0.001$.
In both comparisons, the null hypothesis is that the mean difference between the mean kappa value of the proposed method and each conventional method is zero.

\begin{table}[t]
    \centering
    \caption{\small Performance comparison of CSP, FBCSP and proposed SISAE in terms of mean Kappa value}
    \begin{tabular}{|c|ccc|}
    \hline
    \rule{0pt}{1.em}Test subject & CSP & FBCSP & SISAE       \\
    \hline
   \rule{0pt}{1.2em}Subject 1 &0.259 &0.158 & \textbf{0.717} \\[0.2em]
                    Subject 2 &0.047 &0.062 & \textbf{0.292} \\[0.2em]
                    Subject 3 &0.410 &0.323 & \textbf{0.756} \\[0.2em]
                    Subject 4 &0.331 &\textbf{0.342} & 0.311 \\[0.2em]
                    Subject 5 &0.030 &0.027 &\textbf{0.293}  \\[0.2em]
                    Subject 6 &0.116 &0.059 &\textbf{0.251}  \\[0.2em]
                    Subject 7 &0.063 &0.045 &\textbf{0.388}  \\[0.2em]
                    Subject 8 &0.550 &0.535 & \textbf{0.882} \\[0.2em]
                    Subject 9 &0.225 &0.412 & \textbf{0.614} \\[0.2em]
                    Avg.      &0.226 & 0.218&  \textbf{0.500}\\[0.2em]
   \hline
\end{tabular}
    \label{tab:results}
\end{table}

\section{Conclusion}\label{conclusion}

In this article, we presented a subject-independent framework based on a supervised autoencoder in order to skip the calibration procedure required for new subjects.
The proposed network balanced extracting features ideal for separating MI signals and finding underlying patterns suitable for subject-to-subject generalization.
We evaluated our method on dataset 2a from BCI competition IV.
The simulation results showed that the suggested framework significantly outperformed conventional and widely used CSP and FBCSP algorithms with a p-value less than 0.001.

\bibliographystyle{IEEEtran}
\bibliography{./root.bib}

\end{document}